\documentstyle{mn}
\begin{document}

\input ./BoxedEPS.inp
\SetEPSFDirectory{./}
\SetRokickiEPSFSpecial
\HideDisplacementBoxes


\title[Radio--Optical Correlation in SSQs]{
The radio--optical correlation in steep-spectrum quasars}
\author[Serjeant et al.]
{Stephen Serjeant$^{1,2}$, Steve Rawlings$^{2}$,
Stephen J. Maddox$^{3}$, \vspace*{0.2cm}\\
{\LARGE
Joanne C. Baker$^{4}$,
Dave Clements$^{5}$, Mark
Lacy$^{2}$, Per B. Lilje$^{6}$}\vspace*{0.2cm}\\
$^{1}$Astrophysics Dept., Blackett Labs., Imperial College London,
Prince Consort Road, London SW7 2BZ\\
$^{2}$Astrophysics, Department of Physics, University of
Oxford, Keble Road, Oxford OX1 3RH\\
$^{3}$Royal Greenwich Observatory, Madingley Road, Cambridge, CB3 0EZ\\
$^{4}$Mullard Radio Astronomy Observatory, Cavendish Laboratory,
Madingley Road, Cambridge CB3 0HE. \\
$^{5}$European Southern Observatory, Karl-Schwarzschild-Strasse 2,
D-85748 Garching-bei-Munchen, Germany.\\
$^{6}$Institute of Theoretical Astrophysics,
University of Oslo,  P.O. Box 1029 Blindern, N-0315 Oslo, Norway\\
}

\maketitle
\begin{abstract}
Using complete samples of steep-spectrum quasars, we 
present evidence for a correlation between 
radio and optical luminosity which is not caused by
selection effects, nor caused by an orientation dependence 
(such as relativistic beaming), nor a byproduct of cosmic
evolution. 
We argue that this rules out models of jet formation
in which there are no parameters in common with the production of the
optical continuum. This is arguably the 
most direct evidence to date for a close link between accretion onto a
black hole and the fuelling 
of relativistic jets. The correlation also provides a natural
explanation for the presence of aligned optical/radio structures in only
the most radio luminous high-redshift galaxies.

\end{abstract}

\section{Introduction}

It is now established 
that quasars can be divided into two
physically distinct classes: 
radio-loud quasars (RLQs) and radio-quiet quasars (RQQs) (Peacock,
Miller \& Longair 1986;  
Kellermann {\it et al.} 1989; Miller, Peacock \& Mead 1991; Miller,
Rawlings \& Saunders 1993; Wilson \& Colbert 1995). 
Relativistic jets are a universal feature of RLQs
(e.g. Bridle {\it et al.} 1994) and are also probably associated with at least 
some RQQs (e.g. Miller {\it et al.} 1993); the difference between the 
classes is thus not whether the quasars can form relativistic jets, but
related instead to the fraction of the total power output 
channelled along them in a bulk kinetic form (see also Rawlings 1994).

It is now also established 
that at least some radio galaxies 
harbour obscured quasar nuclei
(e.g. Antonucci 1993;
Antonucci, Hurt \& Kinney 1994; Dey \& Spinrad 1996, Ogle {\it et al.}
1997). However,
the popular notion that the probability of
obscuration is a strong function of the angle between the jet axis
and the line-of-sight (e.g. Barthel 1989; Antonucci 1993) 
due to the ostensible presence of a
dusty molecular torus, is still debated, and may apply
only to a restricted range of radio luminosity and/or redshift
(see e.g. Lawrence 1991; Jackson \& Rawlings
1997 and refs. therein). 
Accepting this orientation-based unification
scheme, however, it is possible to combine RLQs and at least some
radio galaxies into a single `radio-loud' category, and estimate
the power in the photoionising ultraviolet continuum $Q_{\rm phot}$
(which is hidden in the case of radio galaxies) from the
luminosity of the narrow emission lines. By doing this in the 3CRR
sample (Laing, Riley \& Longair 1983) 
Rawlings \& Saunders (1991) inferred that `radio-loud' objects
have bulk powers $Q_{\rm bulk}$ in their jets of the same order 
as $Q_{\rm phot}$, whereas in the case of RQQs (and their possible obscured 
counterparts; see Lonsdale, Smith \& Lonsdale 1989) 
the fraction of the power output channeled into jets is 
$\stackrel{>}{_\sim} 10^{3}$ times lower 
(Miller {\it et al.} 1993; Rawlings 1994). The strongly differing fractions
seem to be a fundamental 
difference between radio-loud and radio-quiet active galactic
nuclei.

Despite this difference
there is some evidence that {\it within} each class of quasar 
there is some intrinsic connection between $Q_{\rm bulk}$ and $Q_{\rm
phot}$. If one accepts that radio luminosity ($L_{\rm rad}$) and 
[O {\sc iii}] line luminosity ($L_{\rm [O~III]}$) are crude indicators
of these 
variables then the separate, but roughly parallel, loci of the
radio-loud and radio-quiet objects in the $L_{\rm rad}$ {\it versus}
$L_{\rm [O~III]}$ plane (e.g. Rawlings 1994) hints at a correlation
between $Q_{\rm bulk}$ and $Q_{\rm phot}$ whether the jet is in a 
radio-loud quasar (with $Q_{\rm bulk} \sim Q_{\rm phot}$) or a radio-quiet
quasar (with $Q_{\rm bulk} \ll Q_{\rm phot}$). 
This gives rise to the hope that these relations
might constrain models for the fuelling of relativistic jets 
in all types of active galaxies 
(e.g. Rawlings \& Saunders 1991; Falcke {\it et al.} 1995).

However, this is far from being a universally accepted view. 
In both radio-loud and radio-quiet objects, the correlation between
$Q_{\rm bulk}$ and the inferred $Q_{\rm phot}$ can be influenced, or
even caused by selection effects in samples used to date, as can the
apparently similar magnitudes of $Q_{\rm bulk}$ and $Q_{\rm phot}$ in
radio-loud objects. We will discuss this in more detail in section
3.4. 

There are further objections, on physical rather than statistical
grounds. 
For example, Dunlop \& Peacock (1993) 
have suggested that $L_{\rm rad}$--$L_{\rm [O~III]}$ correlations 
in radio-loud objects 
are more plausibly explained by models in which 
both luminosities are enhanced 
through interactions between radio jets (and lobes) and a dense environment
(see also Baum, Zirbel \& O'Dea 1995 and refs. therein). 
There are certainly some cases of both radio-loud objects
(e.g. Lacy \& Rawlings 1994) and radio-quiet objects 
(e.g. Axon {\it et al.} 1989) where at least some of 
$L_{\rm [O~III]}$ may be attributable to power supplied by the jet.
A key question, therefore, is whether one can find a direct link 
between $Q_{\rm bulk}$ and $Q_{\rm phot}$ rather than inferring the latter
indirectly from $L_{\rm [O~III]}$.

The optical continuua and nuclear emission lines of RLQs are 
the direct probes of $Q_{phot}$ required.
Dunlop \& Peacock (1993) cite the lack of a correlation between the 
optical continuum and radio luminosities of RLQs as an argument 
in favour of their environmental interpretation of 
the $L_{\rm rad}$--$L_{\rm [O~III]}$ correlation.
In fact the existing evidence on the radio--optical correlation for RLQs 
is mixed (Peacock, Miller \& Longair 1986; 
Browne \& Murphy 1987; Neff, Hutchings \& Gower 
1989; Miller, Peacock \& Mead 1990; Miller {\it et al.} 1993 - we will
review these contradictory claims in section 3.4) and,
in practice, confused by a number of issues. 

Firstly,
Peacock, Miller \& Longair (1986) showed that one must not mix
RLQs and RQQs since doing so can produce an apparently universal (but
spurious) radio--optical correlation. The authors argued that
RLQs have a minimum optical luminosity to explain why 3C
quasars are among the brightest optically, and why the RLQs
in the optically-selected Bright Quasar Survey (Schmidt \&
Green 1983) are among the most luminous radio sources. This alone
would yield a spurious correlation when comparing RLQs with RQQs over
a wide optical range. (In fact our new data rules out this scenario,
as we discuss in section 3.4.)

Secondly, the strong orientation dependence of the optical 
and radio continuua in flat-spectrum core-dominated RLQs 
(e.g. Jackson {\it et al.} 1989; Baker 1996), can lead to
a radio--optical correlation. 
This follows because radio and optical luminosities 
can be simultaneously enhanced in these objects
by relativistic beaming of synchrotron radiation. 

Thirdly, although 
there are hints of coupled optical--X-ray--radio processes in ostensibly 
unbeamed RLQs (Browne \& Wright 1985; Browne \& Murphy 1987), the
samples used to date either lack the spectroscopic redshifts which
would distinguish radio luminosity dependence from evolution, or have
serious and unquantifiable selection biases. 

In this paper we present the results of a new study of the 
radio--optical correlation for quasars in which we have attempted to
implement lessons learnt from the studies referenced above.
We confined our attention to one of the distinct classes of 
quasars, the RLQs, since we believe their radio luminosities are a 
more straightforward indicator of $Q_{\rm bulk}$ (which, 
with suitable radio data, 
it is possible to estimate in individual quasars; 
Rawlings \& Saunders 1991; Rawlings 1993). A recent study of the
radio--optical correlation in RQQs can be found in Lonsdale {\it et
al.} 1995. 
To reduce the effects of relativistic beaming and 
optical continuum anisotropy we have also chosen to concentrate 
on steep-spectrum RLQs (hereafter SSQs), to the exclusion of flat-spectrum,
core-dominated quasars. Section 2 outlines the 
selection and study of our new complete sample of SSQs. 
To distinguish between 
$L_{\rm rad}$-dependence and evolution ($z$ dependence), this sample
is combined with two other complete radio-selected samples which
together span a wide range in $L_{\rm rad}$ at a given redshift; we also
make a comparison with the SSQs in an optically-selected sample.
In Section 3 we present evidence for a radio--optical correlation, and
assess the role of sample selection. In Section 4 we attempt
a physical interpretation of the radio--optical correlation, and 
comment briefly on the implications of this result for studies
of radio galaxies. Throughout this paper, we take values of
\mbox{$H_0=100$ kms$^{-1}$Mpc$^{-1}$}, \mbox{$\Omega=1$} and a
zero cosmological constant.

\begin{figure*}
\hspace*{0cm}
\vspace*{-3cm}
\centering
  \ForceWidth{4.5in}
  \vSlide{-5cm}
  \hSlide{-13cm}
  \BoxedEPSF{}
\parbox{140mm}{
\small
\vspace*{-3cm}
{\bf Figure~1:} The radio luminosity ($L_{408}$), redshift ($z$)
plane for SSQs from the MAQS
(squares), the MQS (stars), 3CRR (circles) and the BQS (crosses). 
The brightest $33\%$ (excluding the BQS) have been plotted as open
symbols. For 3CRR the $178$MHz-$750$MHz spectral indices from Laing,
Riley \& Longair (1983) were used. The loci of
sources with \mbox{$\alpha=0.85$} and  
\mbox{$S_{408}=0.95$} (MAQS/MQS limit) and
$4.94$ Jy (3CRR limit) are shown as dashed lines.
}
\end{figure*}

\section{Molonglo-APM Quasar Survey}

Our new, complete sample of SSQs was drawn initially from the $408$-MHz
Molonglo Reference Catalogue (MRC, Large {\it et al.} 1981), down to a flux
density \mbox{$S_{408}\ge0.95$ Jy} with a radio spectral index
cutoff $\alpha\ge0.5$ (where $\alpha=-{\rm d}\log S_\nu/{\rm d}\log\nu$
evaluated near 1~GHz).
The MAQS survey area is limited to  a \mbox{$\sim1$}-sr region in which
both UK Schmidt APM (Automated Plate-measuring Machine) data and 
Texas Catalogue positions (Douglas {\it et al.} 1996) were available, 
bounded roughly
by $-35^\circ \stackrel{<}{_\sim} \delta \stackrel{<}{_\sim} 0^\circ$,
$21^{\rm h} \stackrel{<}{_\sim} \alpha \stackrel{<}{_\sim} 5^{\rm h}$. 
Hence, the sample is called the Molonglo/APM Quasar Survey
(MAQS; Serjeant 1996, Maddox
{\it et al.} in preparation, Serjeant {\it et al.} in preparation). 
Optical identifications were made from the 
UK Schmidt $b_{\rm J}$ plates, using APM classifications to exclude
``galaxy'' identifications brighter than \mbox{$b_{\rm J} = 20.5$}. 
The quasar candidates therefore consisted of all ``stellar'' or
``merger'' identifications, and all ``galaxy'' identifications fainter
than \mbox{$b_{\rm J} = 20.5$}.
Redshifts have been measured 
(at the Anglo-Australian, William Herschel and Nordic Optical Telescopes) for
all quasar candidates to the plate limit of
\mbox{$b_{\rm J}\simeq22.5$}, yielding a complete spectroscopic
catalogue of $159$ (steep- and flat-spectrum)
confirmed quasars. Optical spectra of all
quasar candidates (including those rejected for lacking broad
optical-UV lines) will appear in a forthcoming paper. 

The deep plate limit is very important to this study: 
only then is the survey likely to detect essentially
all the SSQs as far as $z\simeq 3.3$ where the
Ly$\alpha$ line leaves the $b_{\rm J}$ band. A recent spectroscopic
study of a complete and radio-fainter sample of 7C radio sources 
(Willott et al. 1997) quantifies 
this by finding no SSQs optically fainter than the plate limit used for the
MAQS ({\it i.e.} SSQs with $b_{\rm J}>22.5$ comprise less than $5$ per
cent of the total population), but that most are  
fainter than the $B\approx 20$ limit of studies based, for example, on
POSS-I.

We have combined data on our new sample with data on SSQs from the
radio-selected 3CRR sample (Laing, Riley \& Longair 1983).
In addition, we augment the MAQS sample with SSQs taken from the 
Molonglo Quasar Sample (MQS, Baker 1994), also selected from the MRC
down to $S_{408}=0.95$~Jy, and to the $b_{\rm J}$ plate limit. 
Finally, we include the SSQs from
the optically selected Bright Quasar Survey (hereafter BQS, 
Schmidt \& Green 1983). 
Figure~1 shows $408$-MHz radio luminosity as a function of
redshift ($L_{408}$, $z$) for the four samples; the MQS points 
are plotted in separate symbols where the identification is not shared by
the MAQS. Taken together, these four samples 
provide a wide dispersion in $L_{408}$ at any redshift $z$,
allowing us to distinguish cosmic evolution ($z$ dependence) from
radio luminosity dependence.

We have split the
combined sample into two flux density bins; open symbols are used for the 
$33\%$ of sources with \mbox{$S_{408}>3.23$} Jy, and filled symbols
for the remainder. This split was chosen to yield a roughly even
fraction in the brighter bin throughout the redshift range. Measured
values of radio spectral index, $\alpha$ were used where 
available ($100\%$ of 3CRR and BQS, $97\%$ of MQS,
$72\%$ of MAQS), and \mbox{$\alpha=0.85$} assumed otherwise.
Some of the brighter SSQs, and nearly all the flat spectrum quasars,
should appear in the Parkes catalogue (Wright {\it et al.} 1991): 
for a source on the MAQS radio flux limit the Parkes catalogue
should detect all the radio sources with 
\mbox{$\alpha\stackrel{<}{_\sim}0.4$}. We are thus 
confident that flat-spectrum sources have been almost 
entirely excluded, despite our currently incomplete $\alpha$ information. 

\section{Results}

\subsection{The radio--optical correlation}

\begin{figure}
\centering
  \ForceWidth{3.0in}
  \TrimTop{9cm}
  \hSlide{-9cm}
  \BoxedEPSF{}
\parbox{70mm}{{\bf Figure~3:}
The variation of apparent magnitude with redshift for a quasar of
fixed absolute magnitude. The two curves show the variation for a
numerical integration of the composite quasar spectrum, and the
corresponding variation with a power law optical 
quasar continuum (as discussed in the text). The power law model is
the brighter at a redshift of $z=5$. Both models have been normalised
to produce the same apparent magnitude at a redshift of $1.5$. 
}
\end{figure}

In Figure~2, optical luminosity is shown as a function of $z$, again split
into bright and faint radio subsamples. 
In order to calculate the absolute B-magnitude, $M_{\rm B}$, 
we transformed $b_{\rm J}$ magnitudes to $B$ magnitudes using 
\mbox{$B \approx b_{\rm J}-0.14$} and adopted typical optical spectral indices 
of $0.5$ (the value derived {\it e.g.} from the composite 7C SSQ
spectrum by Willott {\it et al.} 1997). A more careful treatment of
K-corrections was not possible since a significant fraction of the
objects are from the literature, having redshifts but not 
spectrophotometry. 
Typical uncertainties in apparent magnitudes are \mbox{$\sim
\pm 0.2$}~mag; similar contributions to the absolute photometric errors
are made by the uncertainties in the K-corrections.  (Error bars are   
omitted from the plots for clarity.) 

Two possible further sources of error are galactic reddening and
emission line contributions to the $b_{\rm J}$ flux (intrinsic
reddening will be discussed separately below).  The galactic reddening
is typically $A_{\rm B}\simeq 0.2$ magnitudes, and the variations in
galactic reddening are expected
to cause an additional relative photometric error of 
$\sim0.1$ mags (de Vaucoleurs \& Buta 1983a, b). 
This variation is smaller than our typical photometric errors, so we
do not correct for it here. The MAQS extends to galactic latitudes
only as low as $\mid b\mid =30^\circ$, so any galactic reddening is likely
to be more significant for 3CRR; this would only increase the
significance of our results discussed below. Unfortunately not all
the quasars from the literature have published 
line fluxes, so it is not possible to apply empirical emission line
corrections to the $b_{\rm J}$ fluxes uniformly over the sample. 
However, it is well-established in radio quiet quasar
surveys that emission lines make a small contribution to the overall
photometric errors, {\it e.g.} Schmidt \& Green (1983). Since the
$b_{\rm J}$ passband is probably wider 
than the B selection passband of Schmidt \& Green (1983), we might
expect the line effects to be even 
smaller. This is verified in Figure~3, where we have integrated
the composite quasar spectrum of Francis {\it et al.} (1991),
appropriately redshifted, over the $b_{\rm J}$ passband. Also plotted
is the prediction for an $\alpha_{\rm opt}=0.5$ optical power law,
which is the continuum slope determined in the composite SSQ spectrum
of Willott {\it et al.} (1997). The
deviations from the 
power-law model in Figure~3 are dominated by the slightly atypical
(for SSQs) optical spectral index of the Francis {\it et al.} (1991)
spectrum, except at 
redshifts $z\stackrel{>}{_\sim}3.3$ where Ly$\alpha$ exits the $b_{\rm
J}$ passband. Note the smoothness of the composite quasar curve.

\begin{figure*}
\hspace*{0cm}
\vspace*{-3cm}
\centering
  \ForceWidth{4.5in}
  \vSlide{-5cm}
  \hSlide{-13cm}
  \BoxedEPSF{}
\parbox{140mm}{
\small
\vspace*{-3cm}
{\bf Figure~2:} The optical luminosity ($M_{B}$), redshift ($z$)
plane (symbols as Figure~1). The upper dashed line marks the
BQS magnitude limit, the lower line the MAQS limit, and the
central line indicates the approximate threshold of reliable
star-galaxy separation. The reddened quasar 3C22 is marked as two
asterisks joined by a vertical line (see text).
}
\end{figure*}

The loci of SSQs at three
apparent magnitudes are shown as dashed lines in Figure~2; the upper
line represents the magnitude limit of the BQS (\mbox{$B=16$}), the
middle line, the \mbox{$b_{\rm J}=20.5$} 
limit above which optically-extended objects were removed from the
MAQS, and the lower line the \mbox{$b_{\rm J}\simeq22.5$} plate limit of
the MAQS and MQS. 
From Figure~2, any incompleteness in the MAQS caused by
rejecting \mbox{$b_{\rm J} < 20.5$} optically extended objects 
is only expected to be present at \mbox{$z<0.4$}. 
Broad-line radio galaxies from 3CRR are also optically extended 
on POSS plates, and have not been included. 
A point-like reddened 3CRR quasar, 3C22, is plotted in Figure~2 
as two asterisks joined by a line, the lower symbol representing the 
measured $M_{B}$ and the upper representing the magnitude
corrected for reddening (see Rawlings {\it et al.} 1995). 

\begin{figure*}
\hspace*{0cm}
\vspace*{-3cm}
\centering
  \ForceWidth{4.5in}
  \vSlide{-5cm}
  \hSlide{-13cm}
  \BoxedEPSF{}
\parbox{140mm}{
\small
\vspace*{-3cm}
{\bf Figure~4:} The flux-flux radio--optical relation for SSQs (symbols as
Figure~1). Note the horizontal offset between the filled (lower) and
open (upper) symbols.
} 
\end{figure*}

A clear tendency is seen 
for the radio-bright quasars at any $z$ to also be brighter optically. 
This trend cannot be explained solely
on the basis of selection effects since, for example, the 
3CRR SSQs have no optical magnitude limit yet are 
brighter on the whole than the MAQS/MQS quasars. 

Comparing on Figure~2
the SSQs of MAQS/MQS with those of 3CRR (restricting both 
to \mbox{$0.4<z<2$} to ensure well matched comparisons),
we find a mean apparent magnitude of \mbox{$18.36\pm0.20$} for
sources above 3.23 Jy (open symbols), but \mbox{$19.22\pm0.14$} for the
remainder; the
null hypothesis of identical distributions is rejected  
at \mbox{$\simeq99\%$} confidence by
both the Kolmogorov-Smirnov and Mann-Whitney tests. 
For the combined samples in the restricted redshift range
\mbox{$0.4<z<2$}, we find
the optical and radio fluxes are significantly correlated
(\mbox{$\gg99.9\%$} confidence using
Spearman's $\rho$); the same is true of the whole sample
({\it i.e.} unrestricted $z$, Figure~4).  This is the first 
evidence that the apparent radio--optical correlation is not
dominated by a redshift effect.


\begin{figure*}
\hspace*{0cm}
\vspace*{-3cm}
\centering
  \ForceWidth{4.5in}
  \vSlide{-5cm}
  \hSlide{-13cm}
  \BoxedEPSF{}
\parbox{140mm}{
\small
\vspace*{-3cm}
{\bf Figure~5:} The radio--optical relation for SSQs (symbols as Figure~1).
}
\end{figure*}

Figure~5 shows the $L_{408}$-$M_{\rm B}$ correlation implicit 
in Figure~2.  
The correlation in Figure~5 is significant at \mbox{$\gg99.9\%$}
confidence (using Spearman's $\rho$), with a best-fit slope of
\mbox{$-2.5\times(0.6\pm0.1)$} (in the $\log_{10}L_{408}$ {\it vs.}
$M_{\rm B}$ plane) and dispersion of \mbox{$\sim1.6$} 
optical magnitudes (using 3CRR and MAQS/MQS). However, this slope and
dispersion may be biased  
estimators of the true values, because the density of points on the
plane depends on the weighting imposed by the bivariate (radio--optical)
luminosity function (LF), and the sampled comoving volume. 

Thus, by comparing the 3CRR and MAQS/MQS samples we find evidence for
the radio-optical correlation which (for the first time) is
independent of cosmic evolution. Is the correlation present in any of
the samples separately? It might be argued that if a correlation only
appears when samples are combined, then it is more suspect than a
correlation from a sample in which selection
effects are uniform across the sources and where fewer corrections
need to be made to inter-compare the sources. In fact
the correlation also present in the MAQS in
isolation (to a similar significance level), but it is perhaps
instructive to show circumstances in which this type of argument
can be misleading. 
Suppose we had two samples of quasars: the first has
$z=1\pm0.1$ and $B=18\pm0.1$; the second has $z=1\pm0.1$ and
$B=22\pm0.1$. It would probably be impossible to demonstrate a
radio-optical correlation with either sample in isolation, but once
combined the correlation would be far easier to detect. 

The  $L_{408}$-$M_{\rm B}$ correlation in Figure~5 is
supported by the radio properties of the optically faintest SSQs
(the band between the lower two dotted lines in Figure~2).
With the exception of reddened quasars like 3C22, which appears well
above this band once corrected for reddening (Rawlings {\it et al.}
1995), this region includes 
{\it no\/} 3CRR SSQs. It is, however, well populated with MAQS/MQS
SSQs. Although reddened examples certainly exist (Baker \& Hunstead
1995, Willott {\it et al.} 1997), 
SSQs are typically reddened much less severely than 3C22 (or are
reddened so severely, $A_{\rm V}\gg 1$, that they will be classified as 
radio galaxies even on the basis of K-band images), 
since 3C22
has no clear broad emission lines in the observer-frame optical. 
This argues 
strongly against $L_{408}$-dependent reddening as a cause of the
radio--optical relation, an explanation which would
in any case require post-shock temperatures finely tuned to the
destruction of dust in, and only in, the most luminous radio
sources. 

This is not to say that reddening does not 
affect the correlation in any way; we only claim that reddening alone 
cannot cause it. For example, it remains possible that reddening may affect the
scatter in the correlation (see Section 4.2), or that the degree of 
reddening may be linked to the optical luminosity {\it e.g. via} 
dust sublimation.



Finally, although our samples have sufficient coverage of the
radio-optical-redshift parameter space to demonstrate a radio-optical
correlation, they are probably not large enough to address any possible
evolution in this correlation. Interestingly, taking the higher
redshifts in isolation the evidence for the correlation is more 
marginal. However, this may 
simply be due to lack of data, and the restricted dynamic range in
radio power at high redshifts\footnote[1]{Such a selection effect 
may also be responsible for the apparent tightening of the $Q_{\rm
phot}$ {\it vs.} $Q_{\rm bulk}$ correlation seen at high-$z$ in 3CRR
({\it e.g.} 
Rawlings \& Saunders 1991).} (see Figure~1). The same high redshift
radio-optical behaviour is also seen 
in numerical simulations (discussed below) supporting this
interpretation. Nevertheless, it remains possible that the correlation
weakens or perhaps fails at high redshifts, implying a different
mechanism for 
jet formation at high ($z\stackrel{>}{_\sim}2$) and low ($z\simeq1$)
redshifts (section 4).

\subsection{BQS Outliers}

The eleven BQS SSQs lie in an atypical region of the correlation. This 
apparent anomaly is also present in the radio--$L_{\rm [O~III]}$ plot
(Rawlings 
1994), so is unlikely to be (for instance) due to photometric errors
in the BQS. There are two points to made about this feature, which we
will support in the next section with numerical simulations. First,
there is a selection effect 
which necessarily over-populates this region: 
the BQS, which covers half the sky, selects the rare quasars with the 
most extreme $M_{B}$ at any $z$, favouring the high $M_B$ side of a 
radio--optical relation with large intrinsic scatter. In contrast,
the deeper MAQS is limited to $\sim 1$ steradian. Second, the BQS SSQs  
are typically \mbox{$\sim10$} times brighter in the radio
than the faintest end of the radio LF, \mbox{$L_{408}\sim10^{24.5} 
{\rm ~W~Hz^{-1}~sr^{-1}}$}, where the comoving space density is 
an order-of-magnitude higher (Dunlop \& Peacock 1990). The lack of
BQS SSQs with fainter radio luminosities is difficult to explain
without either a radio-optical correlation, or a strongly
luminosity-dependent quasar fraction (Lawrence 1991; Jackson \&
Rawlings 1997; Serjeant {\it et al.} in preparation). For the quasar
fraction to explain the deficit of low-$L_{408}$ BQS SSQs, the SSQ
luminosity function would have to be non-monotonic ({\it i.e.} number
density must not be a strictly decreasing function of
luminosity).  
There is no evidence for this within the MAQS (Serjeant
{\it et al.} in preparation), though it is possible that fainter radio
samples may find such a luminosity cut-off ({\it e.g.} Willott {\it et
al.} 1997). This leaves the radio--optical
correlation as the more plausible explanation. 
Furthermore, the fact that four of the BQS SSQs are also 3C
radio sources (although only two meet the more stringent selection
criteria of 3CRR), and are thus among the most luminous radio sources
in the sky, is clear evidence for the radio--optical correlation.
Nevertheless, in section 3.3 we also show that the
the top left hand corner of Figure~5 may be less well sampled for
SSQs than other regions, so the evidence of a radio-optical
correlation in this part of the radio-optical plane is less
compelling; a much more convincing demonstration is found to follow
from the lack of quasars in the bottom right hand corner.

\begin{figure}
\centering
  \ForceWidth{3.5in}
  \TrimTop{9cm}
  \hSlide{-11cm}
  \BoxedEPSF{}
\centering
  \ForceWidth{3.5in}
  \TrimTop{9cm}
  \hSlide{-11cm}
  \BoxedEPSF{}
\parbox{80mm}{
{\bf Figure~6a:} Simulated data without a radio-optical
correlation.}
\end{figure}

\begin{figure}
\centering
  \ForceWidth{3.5in}
  \TrimTop{9cm}
  \hSlide{-10cm}
  \BoxedEPSF{}
\centering
  \ForceWidth{3.5in}
  \TrimTop{9cm}
  \hSlide{-10cm}
  \BoxedEPSF{}
\parbox{80mm}{
{\bf Figure~6b:} Simulated data with a radio-optical correlation}
\end{figure}


\subsection{Simulated radio--optical relations}

%

As the previous discussion of the BQS outliers illustrates, 
the passage of flux limits across the radio-optical plane make a
qualitative understanding of the selection effects on the radio--optical
relation rather difficult. To clarify the situation we made 
numerical simulations of the data. Note that a quantitative comparison
of the 
radio and optical properties of our samples would involve estimating
the bivariate radio-optical luminosity function, which will be
discussed elsewhere (Serjeant {\it et al.}, in preparation); here we
restrict ourselves to reproducing only the gross properties of our
samples. 

In Figure~6 we show the results of Monte-Carlo simulations of the MAQS,
3CRR and BQS samples. Points were sampled randomly from the three
dimensional probability distribution defined by the bivariate
(radio--optical) luminosity function. In Figure~6a, we assume there is no
radio--optical correlation, {\it i.e.} the bivariate LF is  
\begin{equation}
\Phi(L_{408},M_{\rm B},z)\propto\epsilon(M_{\rm B},z)\beta(L_{408},z)
\end{equation}
with the optical LF, $\epsilon$, taken from Boyle {\it et
al.} (1990), and the radio LF, $\beta$, taken from the ``LDE'' model
of Dunlop \& Peacock (1990). We renormalised the Boyle {\it et al.} LF
to unity, {\it i.e.} 
\begin{equation}
\int\epsilon(M_{\rm B},z)dM_{\rm B}=1
\end{equation}
at all $z$; thus we are only using the {\it shape} of the optical LF,
and the cosmic evolution is determined by the radio LF alone. 
Note that the MAQS SSQs in the ($M_{B}$, $z$) plane of
(a) follow the $b_{\rm J}$ plate limit closely, whereas the 3CRR
quasars have no optical flux limit. 
For simulation (b) (Figure~6b) we use
\begin{equation}
\Phi(L_{408},M_{\rm B},z)\propto\beta(L_{408},z)\gamma(L_{408},M_{\rm B})
\end{equation}
with the radio--optical correlation $\gamma$ modelled as Gaussian scatter
of $1.5$ optical magnitudes about the plotted full line, and again
adopting the LDE LF $\beta$
from Dunlop \& Peacock (1990). 
In both simulations we used \mbox{$\alpha =
0.85$} to convert from $2.7$ GHz to $408$ MHz. 

In both (a) and (b) we sampled $35$ points
from 3CRR, $118$ 
from MAQS and $11$ from BQS, integrating $\Phi$ throughout
$24.5<\log_{10}({\rm P/(W~Hz^{-1}~sr^{-1})})<29$,
$-20>M_{\rm B}>-29$. 
Simulation (b) reproduces our data qualitatively, including the
BQS outliers; the quantitative differences (particularly in the BQS
$z$ distributions) are easily explained if the SSQ optical LF
\mbox{$\int\Phi~dL_{408}$} evolves
differently to the total quasar LF ({\it e.g.} La Franca {\it et al.}
1994), or if the SSQ radio LF differs from that of radio galaxies
(Serjeant {\it et al.} in preparation), or perhaps to unknown
selection effects in the BQS itself (Goldschmidt {\it et al.} 1992). 
The qualitative agreement 
was found to be robust to the assumed slope or dispersion in
the radio--optical correlation. Simulation (a), however, is
grossly inconsistent with our data: 
a general problem for models without a radio--optical
correlation is their inability to explain
why SSQs from a bright radio sample ({\it e.g.} 3C) 
are among the brightest optically, and (more marginally) {\it vice
versa} ({\it e.g.} why the BQS contains high-$L_{408}$ SSQs). 

An interesting feature of simulation (b) is the lack of evidence for a
radio-optical correlation at high redshifts, despite the fact that a
correlation is present. In the simulation this is due to the lack of
dynamic range in radio power at high redshifts; the correlation has a
wide intrinsic dispersion and is not detectable without a wide range
in radio power to compensate. A similar
behaviour is seen in our 3CRR/MAQS/MQS combined sample at high
redshift (Figure~2), again where the radio power dynamic range is least
(Figure~1). Also, in both our data and in simulation (b), the 3CRR
quasars lie slightly on the radio-bright side of the radio-opical
correlation. This may be analagous to the BQS outliers (section 3.2):
3CRR selects the radio-brightest quasars at any $z$, favouring the
high $L_{408}$ side of a broad radio-optical correlation.  

Finally, a very clear (perhaps the clearest) visual demonstration of
our correlation is obtained by comparing Figure~5 with the $M_{\rm
B}-L_{408}$ plane of simulation (a) (Figure~6a, lower panel). The 3CRR
SSQs have no optical flux limit, so in the absence of a correlation
there is nothing to prevent them having optical luminosities at the
faintest end of the SSQ optical LF. This is clearly the case in
simulation (a). However, in our data (and in simulation (b)) we find
the 3CRR SSQs have optical luminosities lying on roughly the same
radio-optical relation as MAQS/MQS. This is obviously difficult to
account for unless the radio and optical luminosities are related in
some way. 

In other words, the lack of SSQs in the bottom right hand corner of
Figure~5 is real; this area has been well-sampled for SSQs. 
This is
clearly highly significant, which we can quantify as follows. In 3CRR
there are $31$ SSQs in the range $10^{26}<L_{408}<10^{27.5}$ W
Hz$^{-1}$ sr$^{-1}$, so if we assign absolute magnitudes uniformly
over the range plotted in Figure~5 we find the probability that
the area is empty to be about $10^{-4}$. This probability would be
much smaller still if we were to use a more realistic optical LF. 

In contrast, the top left corner may arguably be less well
sampled. The deficit of optically bright MAQS quasars could easily be
attributed to the optical luminosity function of SSQs with faint radio
luminosities. This leaves the BQS as the only survey which might
adequately sample this region. The BQS does indeed lack the
radio-faint, optically-bright SSQs which would fill this region, in
agreement with what we might expect from the radio-optical
correlation. 
However, the 
lack of these BQS quasars could perhaps be related to the
incompleteness worries in the BQS (Goldschmidt {\it et al.} 1992). 
Several authors ({\it
e.g.}, La Franca {\it et al.} 1994) have also
noted that the total quasar population appears to
have a much higher fraction of SSQs than at higher redshifts. One
suggested 
explanation of this, which may also help explain the apparent
incompleteness in the BQS, 
is that at such
low redshifts the host galaxies may contribute non-negligibly to the
total magnitudes. As a result, the BQS (by eye) stellar selection may
bias the sample with respect to host galaxy properties at these
redshifts in a complicated manner.

\subsection{Comparison with previous results}

At this point is is worth reviewing previous studies in the light of
our correlation, and 
contrasting the much more problematic
selection effects in previous samples. A common problem is the
inability to distinguish evolution from luminosity dependence. For
example, in section 1 we discussed the apparent correlation in 3CRR
between 
bulk kinetic jet power $Q_{\rm bulk}$ and the photoionising radiation
power output $Q_{\rm phot}$ as estimated from the narrow line
luminosity $L_{\rm NLR}$ (Rawlings
\& Saunders 1991). Also, $Q_{\rm bulk}$ and $Q_{\rm phot}$ have
similar orders of magnitude in 3CRR, again suggesting a link. 
However, we will show that the selection criteria of 3CRR could cause
both this and the $Q_{\rm bulk}$-$Q_{\rm phot}$ correlation, if 3CRR
it taken in isolation.

The $Q_{\rm bulk}$ depends strongly on
the total radio luminosity, which in turn correlates strongly with
redshift in 3C. This secondary correlation could ultimately lead to
spurious relationships within 3C. 
For example, suppose there is no intrinsic $Q_{\rm
bulk}$--$Q_{\rm phot}$ correlation (implying that the correlation in
3CRR is due to some selection effect). 
Also, make the reasonable
assumption that the optical luminosity function of radio-loud quasars
evolves roughly as strongly as its radio-quiet counterpart. Then
the narrow-line luminosity $L_{\rm NLR}$ will correlate strongly with
redshift in 3CRR, because $L_{\rm NLR}$ is evolving, and so the $Q_{\rm
bulk}$--$Q_{\rm phot}$ correlation in 3CRR would be due entirely to
their independent evolution. The apparent
similarity of the orders of magnitude 
of $Q_{\rm bulk}$ and $Q_{\rm phot}$ throughout 3CRR may also be due
to this independent 
evolution, and their apparent similarity would not be preserved in
samples of fainter radio luminosity. A similar critique can be made of
the radio-quiet $Q_{\rm bulk}$-$Q_{\rm phot}$ relation, though in this
case the samples are optically selected rather than radio flux
limited.\footnote{Note though that the differing $Q_{\rm bulk}/Q_{\rm
phot}$ ratios in 
radio-quiet and radio-loud quasars is robust.} However, the presence
of a radio-optical correlation in our samples demonstrates for the
first time that the Rawlings \& Saunders (1991) result is not due to
these selection effects, confirming their interpretation, and 
allows us to predict that the $Q_{\rm bulk}-Q_{\rm phot}$ correlation
will be preserved in complete samples with fainter limiting radio flux
density.

If we assume our observed dispersion in the radio-optical correlation
is close to the intrinsic value, then we can also resolve some of the
previous contradictory claims on the existence of the correlation. 

Peacock, Miller \& Longair (1986) ruled out the null hypothesis that the
apparently bimodal 
distribution of radio-loud and radio-quiet quasars is in fact due to a
universal quasar radio-optical 
correlation (section 1). Our data can also immediately rule out the
alternative 
model suggested by Miller, Peacock \& Mead (1990),
that all radio-loud quasars have a minimum optical luminosity of
$M_{\rm B}\simeq-23$. 
Also, on examining their sample selection it becomes clear why this
study failed to detect the correlation. The Miller {\it et al.}
RLQ sample was wisely restricted to a narrow redshift range, which
counters any differential evolution, but unfortunately was limited to
only six SSQs which spanned less than an order of magnitude in both
radio and optical luminosities. Given the very broad dispersion
apparent in our correlation, it hardly surprising that they did not
detect it. Interestingly, although the Peacock {\it et al.} study did
not explicitly address the SSQ radio-optical correlation, they noted
that several of the quasars in their deeper Parkes
subsamples have lower optical luminosities than those in brighter Parkes
subsamples. The samples were however based  
in part on compilations from the V\'{e}ron \& V\'{e}ron
(1983) catalogue, and no attempt was made to separate SSQs from 
flat-spectrum quasars. 

Other previous inhomogeneous compilations also gave evidence for an SSQ
radio-optical correlation. Neff, Hutchings \& Gower (1989) found a
clear radio-optical correlation distinct from evolution effects,
though their sample was selected from 
the literature to fill the $2.7$GHz radio luminosity, redshift plane
as evenly as possible in \mbox{$1<z<2$}. While this avoids the
tendency towards fainter 
fluxes inherent in flux-limited samples, it is not clear if this method
introduces selection effects of its own since the parent sample is
clearly inhomogenous. 
Browne \& Murphy 1987 also found a clear radio-optical correlation in
lobe-dominated quasars, though
their sample was the radio-selected quasars in the
V\'{e}ron \& V\'{e}ron 
(1983) catalogue with published {\it Einstein} X-ray observations,
which the authors emphasised ``is a very heterogenous sample with all
sorts of unknown selection effects.''

On the other hand, Browne \& Wright (1985) had the benefit of 
complete radio flux limited samples, but lacked complete spectroscopic
redshifts. The radio-optical flux-flux correlation present in their
Figure~1 could then easily be explained by differing redshift
distributions, instead of an intrinsic radio-optical correlation. For
instance, one might reasonably expect that fainter samples extend to
higher redshifts, so would have correspondingly fainter optical
identifications. Such an interpretation is ruled out explicitly in
section 3.1 above. 

Miller {\it et al.} (1993) presented a study of the radio and optical
properties of $z<0.5$ BQS quasars, though the number SSQs was again
too small to detect our correlation (see also Figure~5). However, we
have also already noted that the
completeness of the BQS has been questioned (Goldschmidt {\it
et al.} 1992). 

In summary, the only previous SSQ samples with redshift-independent
evidence for a 
radio-optical correlation were inhomogeneous compilations, so prone
to unquantified (and probably unquantifiable) selection effects. The
complete SSQ samples, on the 
other hand, were either too small or too limited in radio or optical
dynamic range to detect the correlation we have found.

\section{Discussion}

\subsection{A link between accretion and the fuelling of relativistic
jets}

We argue here that the SSQ radio-optical correlation 
(as predicted {\it e.g.} in the jet formation models of Ferreira \&
Pelletier 1995) suggests a close
link between the formation of the jets and accretion onto the central
black hole. 
Our discussion is 
similar to Rawlings \& Saunders (1991), although
now with the benefit of more direct evidence for a link
between accretion and the fuelling of relativistic jets and without
the selection effect ambiguities. 

The narrow range in equivalent widths of broad emission
features ({\it e.g.} Miller {\it et al.} 1993, Francis {\it et al.}
1993) and continuum variability studies imply that the bulk of the 
continuum is produced on sub-parsec scales, and is most naturally
linked to accretion onto a black hole
({\it e.g.} Begelman, Blandford \& Rees 1984).
Therefore, the SSQ radio-optical link appears to arise on sub-parsec
scales with the optical light produced by accretion, and with the
radio luminosity derived from a centrally-formed jet. 
This confirms the view that the dominant influence on $L_{408}$ is 
$Q$, and not the large scale radio source environment (Rawlings
1993). 
 
It is of course possible that the accretion-jet link is not directly
causal, since both processes could share a close link with a third
parameter. Any correlation of the type shown in Figure~5 is often
dismissed as a ``brighter objects are brighter'' effect. 
However, this scenario necessarily requires the existence of
some links to create the scalings, and in the case of the
radio-optical correlation where both luminosities are generated by
processes within the central parsec, such a link is very likely to be
close.

A radio-optical correlation could also be obtained 
if both $M_{\rm B}$ and $L_{408}$ for individual quasars 
evolve separately, but in the same sense, with time. However, if the 
radio lobes of SSQs are short lived ({\it i.e.} lifetimes \mbox{$\ll
H_0^{-1}$}, {\it e.g.} Alexander \& Leahy 1987), requiring multiple
generations of SSQs, then a conspiracy would have to be preserved
from \mbox{$z\sim3$} to \mbox{$z\sim0.4$} despite strongly
changing physical environments 
({\it e.g.} Yee \& Green 1987; Ellingson \& Green 1991;
Haehnelt \& Rees 1993).

The radio-optical correlation  supports models in which 
SSQ jets are fuelled primarily by accretion on to a black hole
({\it e.g.} Begelman, Blandford \& Rees 1984), perhaps {\it via}
magnetically driven winds ({\it e.g.} Spruit 1996; Ferreira \&
Pelletier 1995). 
In general, if any other parameters dominate the jet mechanism 
(such as disk angular momentum, disk structure, 
or related magnetic fields), 
then our correlation implies that they must also control or be
controlled by the accretion rate. In particular, this allows us to
exclude many classes of models in which SSQ
jets are fuelled primarily 
by black hole spin energy and not accretion energy (Rawlings \&
Saunders 1990; Begelman, Blandford \& Rees 1984). Only if the black
hole spin regulates the accretion flow on (possibly) up to kpc scales
can such 
models be sustained; Begelman (1985) presents a model in which he argues
that local regulation of the accretion rate may be present.

It is also worth remembering that radio-quiet quasars follow their own,
but clearly different, radio--optical correlation (although it has not
yet been shown to be redshift independent),
having far lower jet powers than comparable radio-loud quasars 
(Miller {\it et al.} 1993; Lonsdale {\it et al.} 1995). Therefore,  
at least one of the above parameters differ in radio-quiet and
radio-loud quasars.

\subsection{Scatter in the radio--optical correlation}

There are many possible physical interpretations for the
broad scatter in our correlation. We will list some of the more
obvious candidates here.

First, radio luminosity gives an
imprecise measure of the bulk power in the jets ($Q_{\rm bulk}$)
emanating from 
the central engine since it also depends on the gaseous environment of
the radio source (Rawlings \& Saunders 1991; Ellingson {\it et al.} 1991);  
a plot of $Q$ {\it vs.} $M_{\rm B}$
may give considerably less scatter than Figure~5. 
This would be consistent with the smaller dispersion in the
$Q$-$L_{\rm NLR}$ relation (Rawlings \& Saunders 1991) for 3CRR radio
sources.  Unfortunately, at present we lack the radio data 
necessary to make the conversion from $L_{408}$ to $Q$ for the MRC and
3CRR SSQs; the compact, steep-spectrum minority may move the most in the
conversion to $Q$ because their strong confinement and/or young ages
make the $L_{408}$--$Q$ conversion extremely important. These quasars
may also obey a different radio--optical relation if they have
significantly fainter or redder continuua than the main population
({\it e.g.} Baker \& Hunstead 1995). 


Second, reddening of the optical quasar light is very likely
contributing to the scatter (Baker \& Hunstead 1995; Baker 1996). 
It may be possible to reduce
this cause of scatter by deducing reddening from the optical spectra,
but since this typically needs 
higher-quality spectra and/or near-infrared photometry for our MAQS sample
we have not yet been able to investigate this. 

Third, optical variability may also increase the observed scatter and may
also depend on luminosity. However, the variations are expected to be
small for SSQs, about 0.2~mag on average on timescales of years (Hook
{\it et al.}
1994). Variability scatter would be reduced by using median 
magnitudes over a longer timescale. The intrinsic spread in spectral
energy distributions over the observed optical waveband must also
contribute to the scatter (e.g. Elvis 1994). 

Fourth, the sampling of the
radio--optical plane is strongly non-uniform, as discussed above, which
may be responsible for some of the apparent dispersion. 

Last, the dispersion could reflect additional dependences 
on secondary parameters. If this final suggestion is correct, then
other observable quantities in SSQs may correlate with their deviation
from the best fit radio--optical correlation. Such a discovery may 
provide far stronger observational constraints on the formation of the
jets.

\subsection{The optical properties of radio galaxies}

The radio--optical correlation we have established for SSQs 
has important implications  for the orientation-based unified schemes for 
active galaxies ({\it e.g.} Barthel 1989; Antonucci 1993). 
The radio--optical relation suggests that amongst objects
with a naked quasar nucleus those with the most luminous radio 
sources are necessarily also highly luminous in the optical.
If the compact quasar nucleus is hidden from our direct
view, as is now known to be the case in some radio galaxies 
(e.g. Dey \& Spinrad 1996), then in the context of the
orientation-based unified schemes the radio--optical relation
predicts that a high narrow-line luminosity $L_{\rm NLR}$ is
inescapable. The radio--optical relation is then an obvious
candidate for the cause of the relation between narrow-line luminosity 
$L_{\rm NLR}$ and $L_{408}$ for radio sources 
(Baum \& Heckman 1989; Rawlings \& Saunders
1991; McCarthy 1993; Rawlings 1993; Jackson \& Rawlings 1997). 

The SSQ radio--optical relation may also underpin many anomalous 
properties of the most luminous high-$z$ radio galaxies.
For example, most models for the alignment of the
radio and optical structures of high-$z$ radio galaxies
({\it e.g.} McCarthy 1993) require that roughly constant fractions
of $Q_{bulk}$ and $M_{B}$ are used to supply the different aligned
components. If we adopt the view that $Q$ and
$M_{\rm B}$ are closely linked, then as $L_{408}$ drops so 
must the level of aligned light, whether it is supplied
by $Q_{bulk}$ from the jets ({\it e.g.} Lacy \& Rawlings 1994) or by $M_{B}$
from the quasar ({\it e.g.} by scattering; Tadhunter {\it et al.} 1992). 
The lack of
alignments and the low fraction of scattered optical/UV light 
in less-luminous radio sources (Dunlop \& Peacock 1993; Cimatti \& di
Serego Alighieri 1995), independent of redshift, support this
prediction.

\section{Conclusions}

Using complete samples of SSQs, we have shown that
the radio and optical luminosities of SSQs correlate with $\gg99\%$
significance, independent of survey selection and cosmic epoch. 
Such a correlation would naturally explain the observations 
of aligned optical emission in only the most luminous radio galaxies.
Also, we infer that the radio
jets of SSQs are regulated by at least one parameter which is shared 
with the production of the optical continuum; in the accepted standard model
for active galactic nuclei this implies a link 
between accretion and the fuelling of the relativistic jets. 

This is not the first observational suggestion of such a link, nor the
first claim of a radio-optical correlation. However, this is the first
evidence that neither the correlation nor the apparent radio-optical
links are caused by selection effects, particularly those inherent in
single flux-limited samples such as 3CR.

\section*{Acknowledgements}
We thank the staff at the Anglo-Australian, William
Herschel and Nordic Optical Telescopes for technical support. The WHT is
operated on the island of La Palma by the Royal Greenwich Observatory in
the Spanish Observatorio del Roque de los Muchachos of the Instituto de
Astrofisica de Canarias. SS thanks PPARC for a studentship, and for
financial support under grant GR/K98728. 
We would also like to thank John Miller and R. Saunders for stimulating
discussions, and the anonymous 
referee for careful reading of the manuscript and several helpful
suggestions.

\end{document}